\documentclass[conference]{IEEEtran}
\usepackage{amssymb}
\usepackage{graphicx}
\usepackage{enumerate}
\usepackage{pifont}

\usepackage{gensymb}
\usepackage{amsmath}
\usepackage{array}
\usepackage{booktabs}
\usepackage{multirow}

\usepackage{rotating}
\usepackage{epstopdf}
\usepackage{threeparttable} 

\usepackage{color}

\usepackage{algorithmicx}
\usepackage[ruled]{algorithm}
\usepackage[noend]{algpseudocode}
\vspace{-0.2cm}%
\alglanguage{pseudocode}
\usepackage{amsthm}

\ifCLASSINFOpdf
\else
\fi

\begin{document}

\title{R$^3 $PUF: A Highly {\bf R}eliable Mem{{\bf R}istive Device based {\bf R}econfigurable PUF}}

	\author{
		\IEEEauthorblockN{Yansong Gao\IEEEauthorrefmark{2} and Damith C.~Ranasinghe\IEEEauthorrefmark{2}}
		\IEEEauthorblockA{\IEEEauthorrefmark{2}Auto-ID Labs, School of Computer Science, The University of Adelaide, SA 5005, Australia
			\\ \{yansong.gao, damith.ranasinghe\}@adelaide.edu.au}
	}
	



\maketitle

\begin{abstract}
We present a memristive device based R$ ^3 $PUF construction achieving highly desired PUF properties, which are not offered by most current PUF designs: (1) High reliability, almost 100\% that is crucial for PUF-based cryptographic key generations, significantly reducing, or even eliminating the expensive overhead of on-chip error correction logic and the associated helper on-chip data storage or off-chip storage and transfer. (2) Reconfigurability, while current PUF designs rarely exhibit such an attractive property. We validate our R$ ^3 $PUF via extensive Monte-Carlo simulations in Cadence based on parameters of real devices. The R$ ^3 $PUF is simple, cost-effective and easy to manage compared to other PUF constructions exhibiting high reliability or reconfigurability. 
None of previous PUF constructions is able to provide both desired high reliability and reconfigurability concurrently.
\end{abstract}


\IEEEpeerreviewmaketitle

\section{Introduction}
Physically Unclonable Functions (PUFs) exploit the static randomness resulting from uncontrollable process variations to extract instance-specific secrets. Unlike assigned storage digital secrets in a memory, the instance-specific secrets arise during the creation of PUF embedding devices. Attributing to inevitable randomness, PUFs cannot be reproducibly forged even by the original manufacturer. 
Various physical randomness sources have been utilized for building PUFs such as gate-delay~\cite{suh2007physical,maiti2009improving}, power-on state of the static random-access memory (SRAM) ~\cite{holcomb2009power}.
Among memory-based PUFs, SRAM PUFs gained the most attention since they are considered 'for free` due to memory availability in almost all commodity products. As the conventional silicon technology is continuously scaling down and approaching its most material and physical extents, there is a growing urge to look for memory elements for nanoelectronic applications. 

The memristive device is one of promising candidates considering its faster speed, higher density, lower power and non-volatility. Moreover, its fabrication is compatible with current CMOS fabrication process. However, in nano region, currently, they suffer serious variations that somehow deteriorates their performance in memory applications. As the device engineers are always attempting to eliminate the variations. Security applications, for example, PUF constructions based on the memristive devices~\cite{rose2013write,koeberl2013memristor,rose2015performance,mathew2015novel,liu2015experimental,gaomrpuf,gao2015memristive}, actually embrace such prevalent physical variations, because truly prevalent randomness implies more entropy when extracting PUF secrets.

The response (output) given the same PUF is highly desired to be stable when it is re-evaluated upon the same challenge (input). This property is referred to as reliability. Besides that, there are other desirable properties such as reconfigurability and large challenge response pairs (CRPs) space---more entropy extracted within a compact area. Reconfigurability of a PUF is the capability of refreshing its CRPs---evolving the PUF itself into a new instance that exhibits different CRP behaviors, which is of great importance in many application scenarios such as updating electronic tokens---electronic tickets, and preventing downgrading software versions by binding software to hardware~\cite{kursawe2009reconfigurable,katzenbeisser2011recyclable,gao2015emerging}. As the other intermediate benefit, the derived key from an rPUF can also be renovated whenever necessary. Moreover, it is indicated that the rPUF is capable of eliminating some potential attacks such as modeling attacks and reverse engineering~\cite{majzoobi2009techniques}. We also note that the reconfigurability helps to mitigate security concerns that malicious contract manufacturers evaluate PUF secrets without authorization before PUF enrollment by a trusted party. In this context, the PUF secrets can be reconfigured before enrollment phase by the trust party, where the reconfigurability makes PUF secrets evaluated by malicious manufactures useless.

Motivated by these targets, we present a memory-based PUF construction, a highly {\bf R}eliable mem{\bf R}istive element based {\bf R}econfigurable PUF (R$ ^3 $PUF), termed as R$ ^3 $PUF by leveraging peculiarities of memristive devices. Compared with memory PUFs based on CMOS, eg., SRAM, R$ ^3 $PUF has higher density that enables a large CRP space within a compact area attributing to the small footprint of memristive devices. Most significantly, R$ ^3 $PUF achieves both high reliability and reconfigurability, while most of current memory-based PUF constructions may achieve either one of them, but not both, at the cost of expensive additional logic. The main contributions of this work are summarized as follows:

{\bf Highly reliable PUF.} We develop R$ ^3 $PUF to generate highly reliable responses. As a consequence, error correction logic is less or even no more required resulting in a lightweight alternative that can be easily plugged into PUF based key generation applications. Based on our simulation results in~Section~\ref{sect:5_disc}, R$ ^3 $PUF achieves reliability of 100\%.

{\bf Reconfigurable PUF.} We endow R$ ^3 $PUF with reconfigurability without incurring additional area overhead, so that its CRP(s) can be refreshed whenever necessary at no additional costs. We emphasize that the reconfiguration of R$ ^3 $PUF is unpredictable and irreversible.

{\bf Evaluation.} We validate R$ ^3 $PUF performance using a public memristive device model guided with experimental data. We detail the rationale of improved R$ ^3 $PUF performance in comparison with previous memristive device based-PUF constructions~\cite{rose2013write,koeberl2013memristor,rose2015performance,mathew2015novel,liu2015experimental,gaomrpuf,gao2015memristive}.

In next section, we present a background on memristive devices and their propoerties,
followed by concisely survey of one memristive device based reliable PUF and two reconfigurable PUF constructions. We also introduce the important memristive device property unnoticed in previous PUF designs. 
In Section~\ref{Sec:3RPUF} we outline R$ ^3 $PUF design along with its operations. Section~\ref{Sec:ResAndDis} evaluates R$ ^3 $PUF performance and compares it with other memristive device based PUFs.
Section~\ref{Sec:Conclusion} concludes the paper.
\section{Memristive Devices based PUFs}\label{Sec:MemBasedPUFs}

\subsection{Resistance Variation Sources}
\begin{figure}
	\centering
	\includegraphics[trim=0 0 0 0,clip,width=0.45\textwidth]{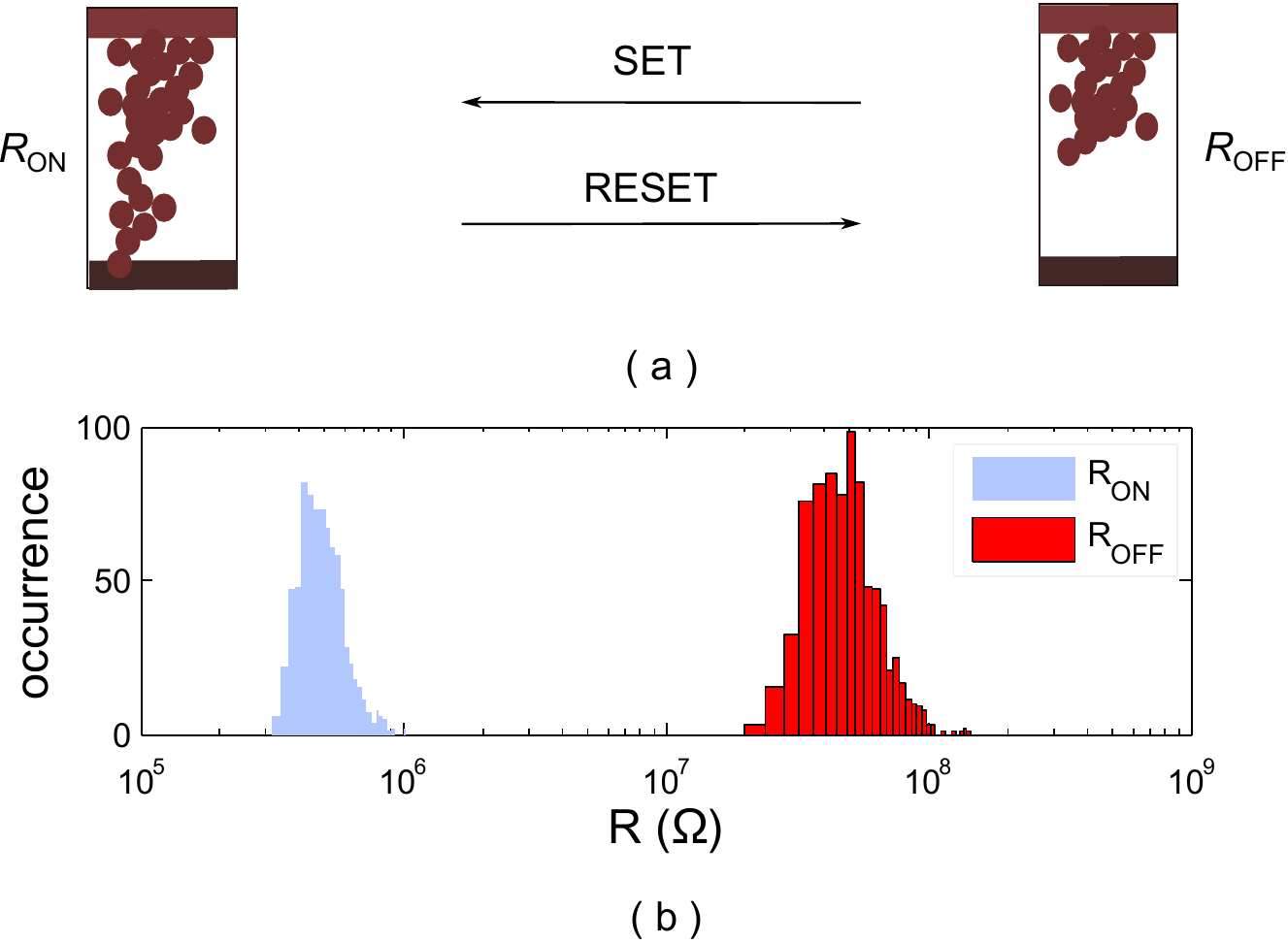}
	\caption{(a) RESET and SET operation of the memristive device. (b) Experimental resistance variation distribution on LRS and HRS state---corresponding to $ R_{\rm ON} $ and $ R_{\rm OFF} $ respectively---measured from 1600 memristive devices~\cite{kim2011functional}.}	\label{fig:memristor}
\end{figure}

A memristive device is a two terminal non-volatile nano element. It switches between high resistance state (HRS) and low resistance state (LRS) by applying a negative/positive potential difference between the top electrode and bottom electrode. The growth and disruption of filamentary conductive paths inside the insulating dielectric are responsible for this switching behavior, illustrated in Fig.~\ref{fig:memristor} (a). The switching from HRS to LRS is referred to as the SET operation, whereas the switching from LRS to HRS is referred to as the RESET operation. The HRS and LRS that are also changeably referred to as $R_{\rm OFF}$ and $R_{\rm ON}$ respectively to denote the two logic states for storing digital information. Attributing to its non-volatility, stored information (resistance) remains after power being cut-off.

The memristive device has a very small footprint. A special layer partially doped is sandwiched between two electrodes. The thickness is down to several nanometers. This enables super high content information storage capability in a compact area where memristive devices are integrated~\cite{wong2015memory}. However, for memristive devices, resistance variations is inevitable and actually prevalent due to the hardness of fine fabrication control~\cite{wong2015memory}. Geometrical variations such as thickness, width, dopant density significantly impact the resistance in both HRS and LRS states, which has been experimentally measured in fabricated devices, see Fig.~\ref{fig:memristor} (b). This is obviously unwanted for memory-based applications because resistance variations deteriorate read margin when distinguishing between HRS and LRS states. However, this prevalent true randomness is of importance for PUF designs~\cite{rose2013write,koeberl2013memristor,rose2015performance,che2014non,chen2015utilizing,mathew2015novel,liu2015experimental,gaomrpuf,gao2015memristive,gao2016physical}.

\begin{figure}[h]
	\centering
	\includegraphics[trim=0 0 0 0,clip,width=0.50\textwidth]{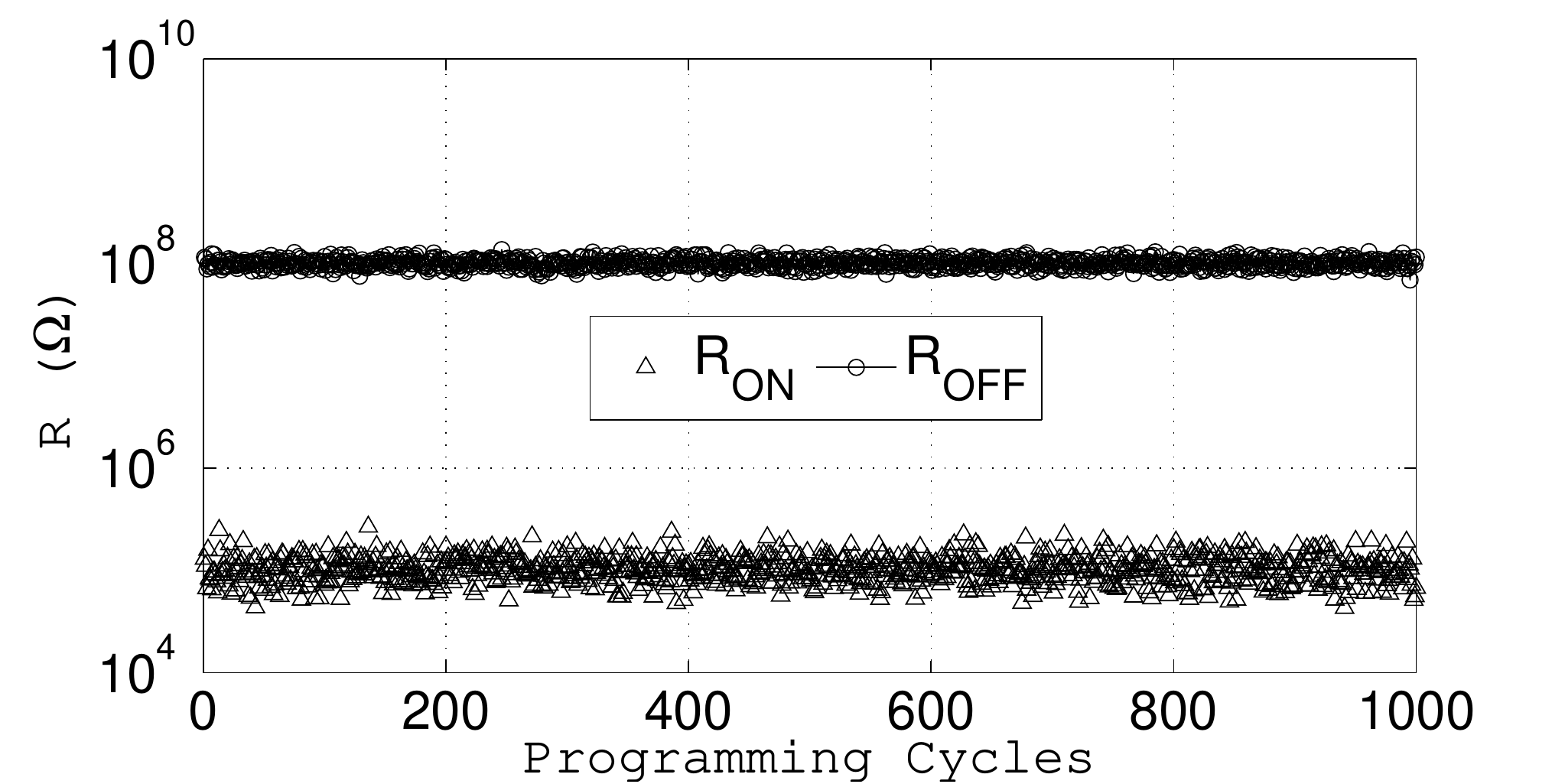}
	\caption{Cycle-to-cycle (C2C) variations. $ R_{\rm OFF} $/HRS and $ R_{\rm ON} $/LRS variation of an individual memristive device for 1000 cycles, experimental data is adopted from \cite{wu2014bipolar}.}
	\label{C2C}
\end{figure}

Besides the geometric induced variations, the resistance in HRS and LRS is also determined by cycle-to-cycle (C2C) variation. C2C variation is an unique variation that is not exhibited by CMOS devices. It is caused by random locations of filaments in the memristive device when it is reprogrammed cycle by cycle---some of these metal filaments' locations are formed and disrupted randomly during reprogramming~\cite{yang2013memristive,wu2014bipolar,chen2015utilizing}. Hence, the resistance observed in HRS or LRS states varies not only among devices but also among different programing cycles given the same device. The C2C variation adopted from measured data is illustrated in Fig.~\ref{C2C}.
\subsection{Reliable PUFs and Reconfigurable PUFs}
\subsubsection{A Reliable PUF} 
\begin{figure}
	\centering
	\includegraphics[trim=0 0 0 0,clip,width=0.45\textwidth]{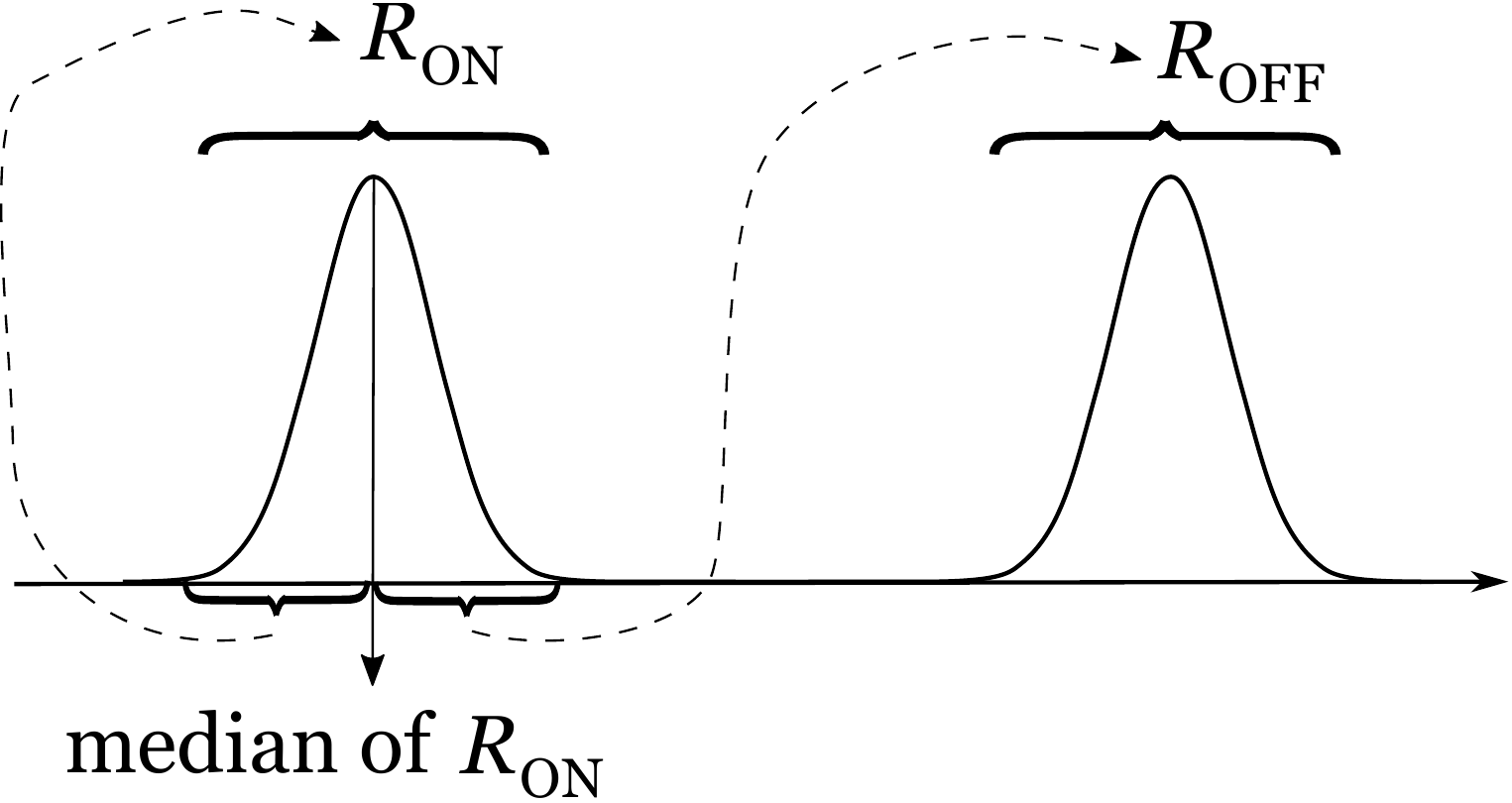}
	\caption{Generalized operations of the memristive device based high reliable PUF in~\cite{che2014non}.}	
	\label{fig:ReliabePUFIlustration}
\end{figure}

Che {\it et al.} \cite{che2014non} proposed a memristive device based PUF that generates highly reliable (error free) response. The PUF operations during enrollment are generalized in Fig.~\ref{fig:ReliabePUFIlustration} and follow steps as: i) All of memristive devices in an array are initially programmed into LRS. This indicates the entropy of PUF response is from the resistance variance in $ R_{\rm ON} $, see Fig.~\ref{fig:memristor} (b). ii) The varied $ R_{\rm ON} $ resistance of each memristive device is digitalized. Then the the median of all digitalized values is determined. iii) The memristive device is programmed into LRS if the digitalized value of such a specific device is lower than the median value, otherwise the memristive device is programmed into HRS.

In general, in Fig.\ref{fig:ReliabePUFIlustration}, if the memristive device initial resistance is lower than the median of LRS ($R _{\rm ON} $), its final state is LRS state. Otherwise, its final state is in HRS state. The regeneration of PUF response is simply reading out the memristive device's state, HRS/LRS is eg., `1'/'0'. The ratio of HRS to LRS is large enough to distinguish these two states, see Fig.~\ref{fig:memristor} (b), in other words, there is no overlap because of a large gap between bimodal resistance distributions. Thereupon, the reliable response regeneration is ensured.

Although this PUF capture certain desirable features from memristive devices such as substantial resistance variations in either HRS or LRS, high resistance ratio of HRS to LRS and non-volatility. The operation procedures during the enrollment is, however, complicated. Such a complicated enrollment phase results in higher cost due to the usage of analog-to-digital converter, counter, and also write-back circuits. As will be shown in the R$ ^3 $PUF, we require neither any extra hardware nor the write-back operation.

\subsubsection{Reconfigurable PUFs}
Gao {\it et al.}~\cite{gao2015memristive} and Chen {\it et al.}\cite{chen2015utilizing} noticed that unique C2C variations of memristive offers the feasibility to design rPUFs. In general, these two memristive device based rPUF constructions exploit either HRS or LRS resistance distributions as random source to extract responses. The response is produced based on comparisons of two (or more) memristive devices' resistance when they are aligned to the same state (HRS/LRS). The response readout uses a small voltage without disturbing the resistance of memristive devices. Whenever reconfiguration is on demand, memristive devices are reprogrammed using a larger voltage that is able to SET/REST the devices.
	
\subsection{Abrupt Switching Behavior} 
\begin{figure}
	\centering
	\includegraphics[trim=0 0 0 0,clip,width=0.45\textwidth]{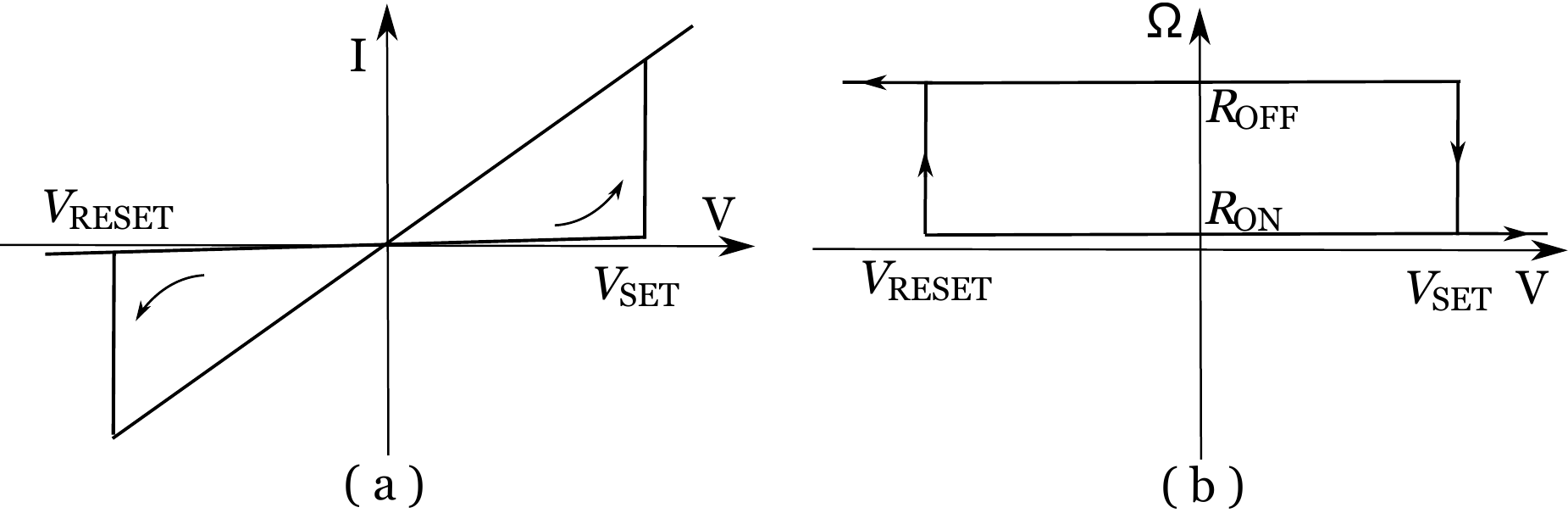}
	\caption{Typical I-V (current-voltage) curve (a) and the corresponding $ \Omega $-V curve (b) of a bipolar memristive device. Both plotted in linear scale,}
	\label{fig:MemristorWithoutRect}
\end{figure}

The abrupt switching behavior reacts with the threshold voltage phenomena~\cite{yang2013memristive}---illustrated in Fig.~\ref{fig:MemristorWithoutRect}. For electric-field-induced bipolar switching in a memristive device, the applied electric field plays a dominant role. A small electric-field is inadequate to move ions in the device to change its resistance~\cite{yang2013memristive}. Therefore, in practice, the resistance of the memristive device stays unchanged or changes negligibly if the voltage drops across the memristive device falling within two threshold voltages $ V_{\rm RESET} $ and $ V_{\rm SET} $. Otherwise, it swiftly switches to HRS if the biased voltage reaches to $ V_{\rm RESET} $ or to LRS if the biased voltage goes to $ V_{\rm SET} $.

By exploiting C2C variations and abrupt switching behavior of memristive devices, our R$ ^3 $PUF achieves high reliability and reconfigurability concurrently with a very simple design and operations.

\section{R$ ^3 $PUF Design}\label{Sec:3RPUF}
In this Section, we detail the principles of R$ ^3 $PUF design by making most of all forgoing memristive device properties.

\subsection{R$ ^3 $PUF construction}\label{Sec:R3PUFConstruction}
\begin{figure}
	\centering
	\includegraphics[trim=0 0 0 0,clip,width=0.40\textwidth]{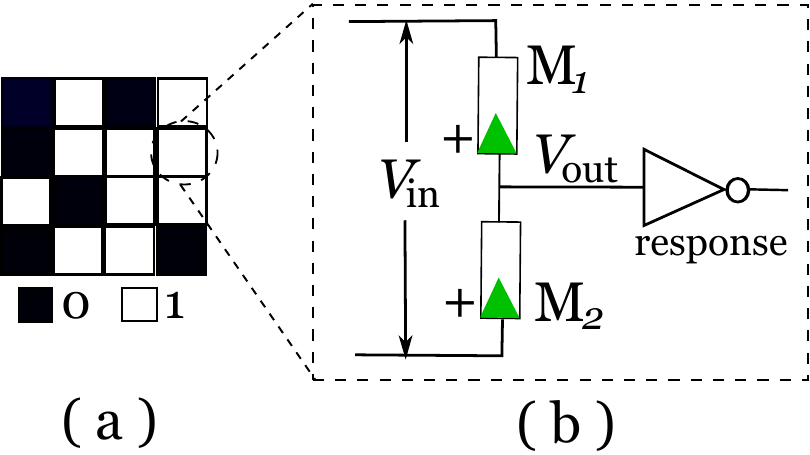}
	\caption{(a) Basic R$ ^3 $PUF cells are placed as an array on chips, alike to memory-based PUFs, address of the R$ ^3 $PUF cell is the challenge. (b) Basic R$ ^3 $PUF cell circuit. Two memristive devices are serially connected. The inverter acts as a voltage comparator to produce a digitalized response.}
	\label{fig:3RPUF}
\end{figure}

Topology of the  R$ ^3 $ PUF is similar to popular memory-based PUFs~\cite{holcomb2009power,zeitouniremanence}. A memory-based PUF consists of a number of memory cells, each cell produces a response bit that is independent on others generated by other cells. The R$ ^3 $PUF, as illustrated in Fig.~\ref{fig:3RPUF} (b), has an alike topology, where the challenge is the address of the R$ ^3 $PUF cell in a high density array. The R$ ^3 $PUF cell is quite simple, where two memristive devices M$_1 $, M$ _2 $ are connected serially. Its operation has two phases: response extraction and readout.

\subsubsection{Response Extraction.} 
\begin{figure}
	\centering
	\includegraphics[trim=0 0 0 0,clip,width=0.35\textwidth]{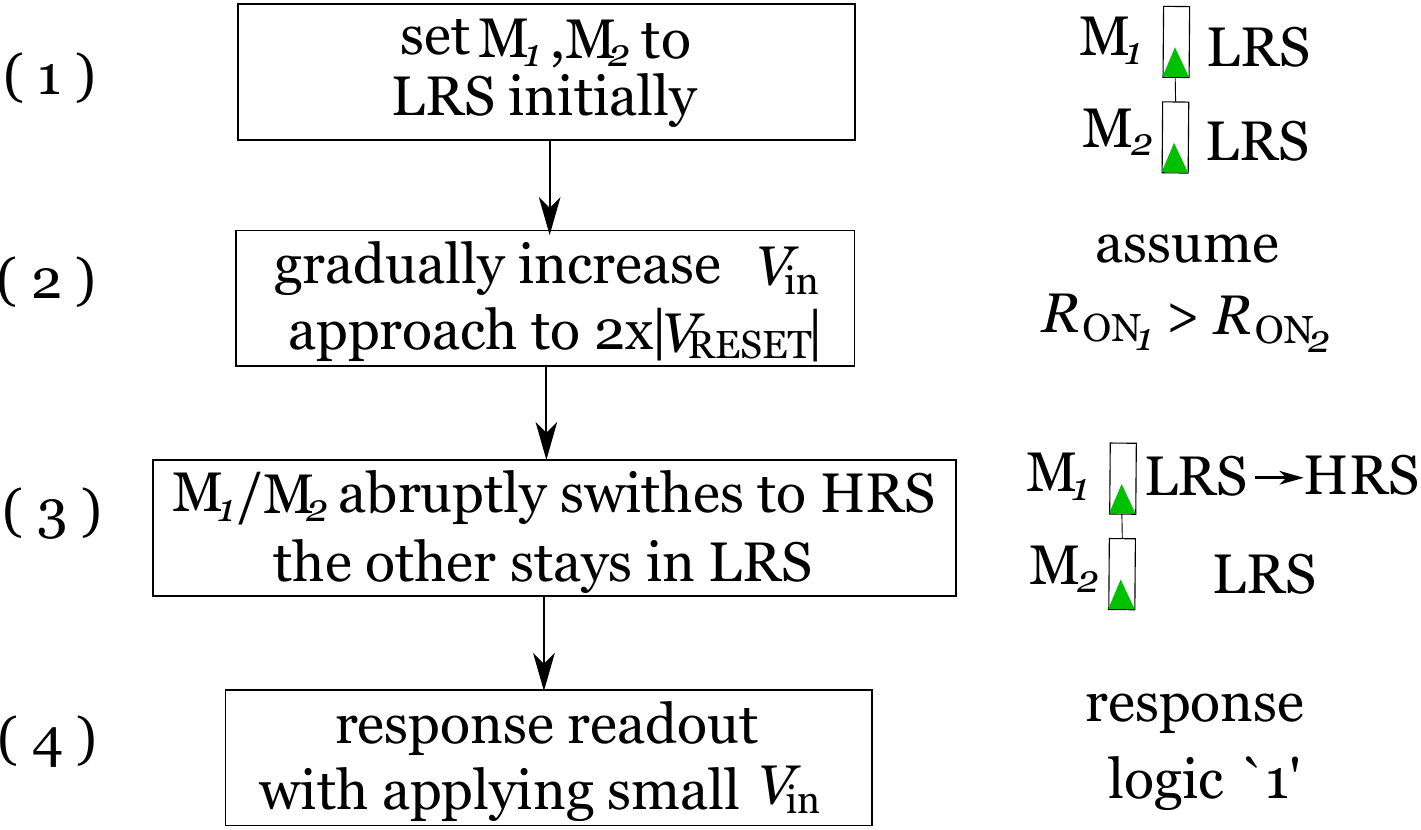}
	\caption{R$ ^3 $ PUF operation procedures, it requires no extra hardware.}
	\label{fig:3RPUFProcedure}
\end{figure}

Both memristive devices are set to LRS/$ R_{\rm ON} $ initially---also see visualized operation procedures depicted in Fig.~\ref{fig:3RPUFProcedure}. Then the applied voltage $ V_{\rm in} $ across two devices gradually increases to $2\times |V_{\rm RESET}|$---noting the polarity of memristive device. Due to inherent resistance variations, $ R_{\rm ON1} $ (LRS resistance of M$_1 $) and $ R_{\rm ON2} $ are different. For the purpose of easing the description, we assume that $ R_{\rm ON1}>R_{\rm ON2} $. As a consequence, M$_1 $ will first reach to the RESET threshold voltage $ V_{\rm RESET} $ and start switching to the $ R_{\rm OFF1} $. This is because M$_1 $ shares more applied voltage considering the fact that M$_1 $ and M$_2 $ eventually forms a voltage divider. Once M$_1 $ starts switching to $ R_{\rm OFF1} $, its increasingly shared voltage further amplifies its switching and makes it switching to $ R_{\rm OFF1} $ even faster. At the meantime, M$_2 $ is stuck in $ R_{\rm ON2} $ because it cannot reach to its threshold voltage as the fact that the voltage dropped across it becomes even smaller once M$_1 $ starts switching. In our R$ ^3 $PUF construction, this amplification alike voltage sharing behavior eventually serves as one basis of the high reliability.

\subsubsection{Response Readout.} The readout is performed by simply applying a small voltage that not disturbs resistance of memristive devices. Recall that at the end of response extraction phase, the M$_1 $ is in $ R_{\rm OFF1} $ and the M$_2 $ is in $ R_{\rm ON2} $.  According to
\begin{equation}\label{Eq:divider}
V_{\rm out}=V_{\rm in}\times \frac{R_{\rm ON2}}{R_{\rm ON2}+{R_{\rm OFF1}}}
\end{equation}
and $ R_{\rm OFF}/R_{\rm ON} $ is usually large---1000 has been experimentally shown in~\cite{kim2011functional}. We can see that the $ V_{\rm out} $ will close to 0~V. We further digitize $ V_{\rm out} $ to obtain a response by simply utilizing an inverter---acting as a voltage comparator. Hence, the response of `1' is produced in this exemplary case. 

Considering M$_1 $ remains in HRS and M$_2 $ in LRS even when the power is off, therefore, whenever the response is regenerated later, it will be stably reproduced. To sum it up, the exploitation of non-volatility of memristive device and high ratio of HRS to LRS enables robust response regeneration.

\subsection{R$ ^3 $PUF Reconfiguration}
To endow the R$ ^3 $PUF with reconfigurability, firstly, the M$_1 $ is reprogrammed back to $ R_{\rm ON1}^{\prime} $ from $ R_{\rm OFF1} $---we follow the same motivating example in Section~\ref{Sec:R3PUFConstruction}---by applying $ V_{\rm in}$ that is smaller than $-V_{\rm SET} $, again noting the polarity of memristive devices in Fig.~\ref{fig:3RPUF}. Now taking the C2C variation into consideration, the $ R_{\rm ON1}^{\prime} $ of M$ _1 $ after this SET operation is different from the previous value. As a result, the relationship between $ R_{\rm ON1}^{\prime} $ and $ R_{\rm ON2} $ becomes unknown---which one is higher is nondeterministic. Next, response extraction operations, step 2 to step 3 depicted in Fig.~\ref{fig:3RPUFProcedure}, are conducted to extract a refreshed response, where the response value is unpredictable.

We realize that the number of times that R$ ^3 $PUFs can be reconfigured is determined by the endurance of memristive devices. Endurance is the maximum cycles that the memristive device can be reprogrammed between HRS and LRS without suffering obvious read margin degradation. Specifically, the ratio of HRS/LRS still keeps higher enough to explicitly distinguish two logic states: `0' and `1'. Luckily, the endurance usually is high, for example, experimental reports of $ 10^5 $ in~\cite{jo20143d}. This number is adequate for most rPUF applications.

\section{R$ ^3 $PUF Evaluations and Discussions}\label{Sec:ResAndDis}
We evaluate and analyze the R$ ^3 $ PUF performance based on the below described public behavior model guided with experimental data as shown in Fig.~\ref{fig:memristor}~\cite{kim2011functional}. Then we fairly compare it with other memristive device based PUFs.
\subsection{Memristive Device Model and Simulation Setup}
\subsubsection{Device Model}
The memristive device model is adopted from \cite{gao2016future}. It has a state variable $ \omega \subset [0,1] $ corresponding to the value of its resistance $ R_{ m} $, which is a function as
\begin{equation}\label{equ:Rm}
R_{ m} =  R_{\rm OFF}(R_{\rm ON}/R_{\rm OFF})^\omega
\end{equation}
According to Eq.~\ref{equ:Rm}, $ R_{ m} = R_{\rm OFF} $ when $ \omega = 0 $, while $ R_{ m} $ is in the $ R_{\rm ON} $ state if $ \omega = 1 $.

The dynamic switching behavior of the memristive device is defined as:
\begin{equation}
{dw\over dt} = \left\{ 
\begin{array}{l l l}
\alpha (v - V_{\rm SET}), & \quad (v\ge V_{\rm SET})\\
\alpha (v + V_{\rm RESET}), & \quad (v\le V_{\rm RESET})\\
\beta v, & \quad \rm otherwise,
\end{array} \right.
\end{equation} 
where $ v $ is the voltage drop through the memristive device. Althoug we consider a symmetrical threshold voltage for $ V_{\rm RESET}$ and $ V_{\rm SET}$, it will not affect the performance evaluation. The $ \alpha $ and $ \beta $ coefficients are switching rates. Here, we set $ \beta = 0 $ assuming that the smaller voltage does not alter the state variable as pointed out in \cite{yang2013memristive}. The $ \alpha=10^5 $ to fit the experimentally reported abruptly switching when $ v>V_{\rm SET} $ or $ v<V_{\rm RESET} $.

The developed behavioral model created in Verilog-A language is also adopted from~\cite{gao2016future}. Simulated I-V curve in Fig.~\ref{fig:SwitchIllustrate} (a) shows well matching behavior as in Fig.~\ref{fig:MemristorWithoutRect}.
\subsubsection{Simulation Setup}\label{Sec:Setup}
The memristive device model is integrated into the R$^3  $PUF cell circuit as shown in Fig.~\ref{fig:3RPUF}, the inverter is implemented by standard 90nm technology. The simulation is conducted via Monte Carlo command in Cadence. Variations for different parameters are listed in Table~\ref{tab:parameters}. Note the variance of $ R_{\rm ON} $ and $ R_{\rm OFF} $ is intentionally set smaller than the measured data in Fig.~\ref{C2C} in order to demonstrate that slightly variation is already able to lead a high reliable R$ ^3 $PUF.

\begin{table}
	\centering 
	\caption{Parameter Settings} 
	\resizebox{0.40\textwidth}{!}{
	\begin{tabular}{c c c c}
		\toprule 
		Parameter  & Mean & Std & Distribution Type\\ 
		\midrule 
		$ R_{\rm ON} $ & $ 5\times 10^{5} \Omega $ & 5\% & Lognormal\\ 
		$ R_{\rm OFF} $ & $ 5\times 10^{8} \Omega $ & 5\% & Lognormal\\ 
		$ V_{\rm RESET} $ & $ - 1 $ V & 5\% & Gaussian\\ 
		$ V_{\rm SET} $ & 1  V & 5\% & Gaussian\\ 
		 transistor length& 120 nm & 5\% & Gaussian\\
		 transistor width& 90 nm & 5\% & Gaussian\\ 
		\bottomrule 
	\end{tabular}}
	\label{tab:parameters} 
\end{table}


\subsection{Results}
\begin{figure*}
	\centering
	\includegraphics[trim=0 0 0 0,clip,width=\textwidth]{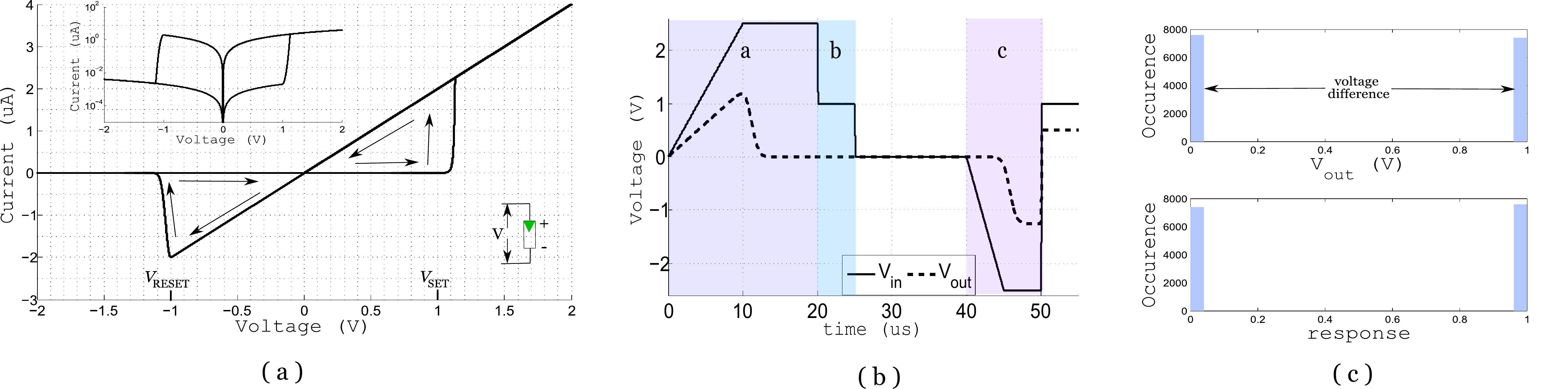}
	\caption{(a) Simulated I-V (current-voltage) curve of a typical memristive device. The inset figure shows the same I-V curve but the y-axis is on a logarithmic scale. $ V_{\rm RESET}=-1 $~V and $ V_{\rm SET}=1 $~V. (b) Validation of the R$ ^3 $PUF operations. (c) Distribution of the $ V_{\rm out}$ and response of the R$ ^3 $PUF.}
	\label{fig:SwitchIllustrate}
\end{figure*}

We carry out simulations according to the R$ ^3 $PUF operations depicted in Fig.~\ref{fig:3RPUFProcedure}. The R$ ^3 $PUF operation is simple that only involves with controlling the applied voltage $ V_{\rm in} $. Therefore, we depict the setting of $ V_{\rm in} $ and the corresponding output voltage $ V_{\rm out} $ between M$ _1 $ and M$ _2 $ in Fig.~\ref{fig:SwitchIllustrate} to experimentally imply the entire operation procedures by an indirectly means. We first describe a specific validation of a single R$ ^3 $PUF cell before large population evaluations. The M$ _1 $ and M$ _2 $ are set to be LRS state initially, where $ R_{\rm ON1}=5\times 10^5~ {\rm \Omega} $ and $ R_{\rm ON2}=4.99\times 10^5~{\rm \Omega} $ with a very small resistance difference of 0.2\% compared with $ R_{\rm ON1} $.

Starts from shadowed area a in Fig.~\ref{fig:SwitchIllustrate} (b), $ V_{\rm in} $ is gradually increased from 0~V to 2.5~V. Both of M$ _1 $ and M$ _2 $ stay unchanged as $ V_{\rm in} << 2 $~V considering that $ V_{\rm out} $ is linear increased as $ V_{\rm in} $ increases. Once the voltage approaches to $ 2~{\rm V} =2\times |V_{\rm RESET}|$, the M$ _1 $ starts switching first as it shares a larger voltage---recall $ R_{\rm ON1} > R_{\rm ON2}$, hence, it reaches to RESET threshold first. The resistance of M$ _1 $ switches from $ R_{\rm ON1} $ to $ R_{\rm OFF1} $ that is amplified by the abrupt switching behavior as the voltage dropped across M$ _1 $ becomes larger. Such a switching will be quickly finished, and later almost all applied voltage drops across M$_ 1 $ according to Equation (\ref{Eq:divider}) and a large HRS/LRS ratio, thereupon, the $ V_{\rm out} $ goes down to near 0~V. The response readout is validated by applying a $ V_{\rm in} $ of 1~V as shown in area b of Fig.~\ref{fig:SwitchIllustrate}. As the $ V_{\rm out}\approx 0 $~V, the response gives logic `1' after digitalization by an inverter. 

Monte Carlo simulation runs 15,000 times taking all the variation sources in Table~\ref{tab:parameters} into consideration. In other words, 15,000 R$ ^3 $PUF cells are simulated. Histograms of $ V_{\rm out} $ and response are shown in Fig.~\ref{fig:SwitchIllustrate} (c). We can see that the $ V_{\rm out} $ is either close to 0~V or 1~V when the readout voltage $ V_{\rm in}=1 $~V is applied. The percentage of `1' in response is 50.60\%, which is close to the ideal value of 50\% that implies good randomness/unpredictability.


\subsection{Discussions}
\label{sect:5_disc}
\subsubsection{Reliability}

Due to the unavailable memristive device temperature model~\cite{gao2015memristive,rose2015performance}, the reliability performance under different temperatures cannot be evaluated. However, the high reliability can be envisioned because of the rationale: i) Once the step 3 of R$ ^3 $PUF operation is done, the resistance of M$ _1 $ and M$ _2 $ will remain constantly even the power is off attributing to the non-volatility. ii) $ V_{\rm out} $ demonstrated in Fig.~\ref{fig:SwitchIllustrate} (c) shows a widely and clearly separation indicating that there is always one memristive device is in LRS state and the other one is in the opposite HRS state. iii) Seeing the high ratio of $ R_{\rm OFF} $ to $ R_{\rm ON} $, even right tail of $ R_{\rm ON} $ distribution and left tail of $ R_{\rm OFF} $ distribution still has magnitude difference. Foregoing facts guarantee that the later re-readout of response is robust. Actually, the histogram of $ V_{\rm out} $ from 15,000 R$ ^3 $PUF cells implies a reliability of 100\%.
\subsubsection{Reconfigurability}
To reconfigure the R$ ^3 $PUF, SET both of M$ _1 $ and M$ _2 $ to LRS is needed. This is validated in the area c in Fig.~\ref{fig:SwitchIllustrate} (b). After SET operation, both memristive devices are in LRS. Considering the physical means induced C2C variations, 
the response generated once the R$ ^3 $PUF is reconfigured cannot be predicted even the previous response value is known. This guarantees the forward security. Likewise, observing the later generated response after reconfiguration is unable to discover the previous response value, which ensures the backward security. Further, the reconfiguration by using C2C variation cannot be reversed by any party.
\subsubsection{Comparisons}
The results of R$ ^3 $PUF are from simulations, note other memristive device based PUF realizations~\cite{rose2013write,koeberl2013memristor,rose2015performance,che2014non,chen2015utilizing,mathew2015novel,liu2015experimental,gaomrpuf,gao2015memristive} also evaluated from simulations. Therefore, it is reasonable to compare with them. 

We are the first work to achieve both high reliability and reconfigurability, moreover, without extra area cost and complicated operations. These works~\cite{rose2013write,koeberl2013memristor,rose2015performance,mathew2015novel} do not take the inherent C2C variations into consideration, which may further degrade these PUF designs' reliability performance. Two works consider the C2C variations to design rPUFs~\cite{chen2015utilizing,gao2015memristive} but without realizing high reliability. Given the only work shows high reliability~\cite{che2014non} reviewed in Section~\ref{Sec:MemBasedPUFs}, it requires complex operation procedures and costs extra hardware overhead. All the aforementioned designs do not fully take advantage of the properties exhibited by the memristive devices, which maybe the reason that they cannot offer high reliability and reconfigurability simultaneously with a very simple circuit implementation. The R$ ^3 $PUF exploits more inherent properties from the memristive devices: process variations, C2C variations, abrupt dynamic switching behavior, high ratio $ R_{\rm OFF} $/$ R_{\rm ON} $, high endurance and the non-volatility. All of these exploited properties of memristive devices eventually lead to a simple R$ ^3 $PUF but equipped with better performance. 

\section{Conclusion}\label{Sec:Conclusion}
In this paper, we propose the R$ ^3 $PUF design and evaluate its performance by extensive simulations guided with existed memristive device model and parameters from experimental data. The R$ ^3 $PUF has higher reliability performance and also is reconfigurable without extra area cost and complicated operations, because we are able to capture peculiarities of memristive devices. Based on the simulation, it indicates that the reliability is almost to 100\%, if not 100\% as infinitely Monte Carlo runs cannot be achieved. Therefore, the ECC implementation overhead can be significantly reduced---or maybe omitted based on the error free response shown in the results---when the R$ ^3 $PUF is used for cryptographic key generations. In addition, The unique C2C variation is exploited to reconfigure the CRPs of the R$ ^3 $PUF, which enables updating the derived cryptographic keys on demand. 

Our future work will experimentally evaluate R$ ^3 $PUF's performance based on fabricated memristive devices.

\end{document}